\documentstyle[prd,aps,floats]{revtex}
\begin{document}
\draft
%%%%%%%%%%%%%%%%%%%%%%%%%%%%%%%%%%%%%%%%%%%%%%%%%%%%%%%%%%%%%%%%%%%%%%
%
%  Uncomment following four lines and one below for 2 column format
%  and figure insertions.
%
\input epsf
\renewcommand{\topfraction}{0.8}
\twocolumn[\hsize\textwidth\columnwidth\hsize\csname
@twocolumnfalse\endcsname
%%%%%%%%%%%%%%%%%%%%%%%%%%%%%%%%%%%%%%%%%%%%%%%%%%%%%%%%%%%%%%%%%%%%%%
\preprint{CERN-TH/97-149, hep-th/9707059}
\title{Dual Inflation}
\author{Juan Garc\'{\i}a-Bellido}
\address{Theory Division, CERN, CH-1211 Geneva 23, Switzerland}
\date{October 1, 1997}
\maketitle
\begin{abstract}
  We propose a new model of inflation based on the soft-breaking of
  N=2 supersymmetric SU(2) Yang-Mills theory. The advantage of such a
  model is the fact that we can write an exact expression for the
  effective scalar potential, including non-perturbative effects,
  which preserves the analyticity and duality properties of the
  Seiberg-Witten solution. We find that the scalar condensate that
  plays the role of the inflaton can drive a long period of
  cosmological expansion, produce the right amount of temperature
  anisotropies in the microwave background, and end inflation when the
  monopole acquires a vacuum expectation value. Duality properties
  relate the weak coupling Higgs region where inflation takes place
  with the strong coupling monopole region, where reheating occurs,
  creating particles corresponding to the light degrees of freedom in
  the true vacuum.
\end{abstract}

\pacs{PACS numbers: 98.80.Cq \hspace{3.6cm} Preprint CERN-TH/97-149, 
hep-th/9707059}

%%%%%%% Comment the next line before submission
\vskip2pc]

The inflationary paradigm~\cite{book} not only provides a very elegant
solution to the classical problems of the hot big bang cosmology, but
also predicts an almost scale invariant spectrum of metric
perturbations which could be responsible for the observed anisotropy
of the cosmic microwave background (CMB). There are a dozen or so
different models of inflation, motivated by distinct particle physics
scenarios. In most of these models one has to make certain
approximations: either they do not include large quantum corrections
to their small parameters or they do not consider possibly strong
non-perturbative effects.

One of the most robust aspects of inflation is the fact that, as long
as there is an almost flat direction, the universe will expand
quasi-exponentially, independently of the nature of the effective
inflaton field. In most models, however, the inflaton is a fundamental
scalar field and the scalar potential is put in by hand, except
perhaps in Starobinsky model, where the field that drives inflation is
related through a conformal transformation to higher derivative terms
in the gravitational action, see Ref.~\cite{book}.

In this letter we propose a new model, based on recent exact results
in duality invariant supersymmetric Yang-Mills theories, where the
inflaton field is a composite scalar field that appears in the theory.
In the last couple of years there has been a true revolution in the
understanding of non-perturbative effects in N=1 and N=2
supersymmetric quantum field theories, following the work of Seiberg
and Witten~\cite{SW} in N=2 supersymmetric SU(2). Duality and
analyticity arguments allow one to write down an {\em exact} effective
action for the light degrees of freedom in both the weak and in the
strong coupling regions. However, our world is non-supersymmetric and
therefore one should consider the soft breaking of N=2 supersymmetric
SU(2) directly down to N=0. This was successfully done in
Ref.~\cite{Luis}, via the introduction of a dilaton spurion superfield
which preserves the analyticity properties of the Seiberg-Witten
solution. The result is a low energy effective scalar potential which
is {\em exact}, i.e. includes all perturbative and non-perturbative
effects. This potential presents an almost flat direction that could
drive inflation. The results of Ref.~\cite{Luis} were studied in the
context of low energy QCD. This letter simply draws the attention to
the possibility that the exact scalar potential found in~\cite{Luis}
could be consistently used at much higher energies, of order the GUT
scale, and be responsible for cosmological inflation. The advantage
with respect to other inflationary models based on N=1 supersymmetry
is the complete control we have on the scalar potential, both along
the quasi-flat direction and in the true vacuum of the theory, where
reheating takes place.

In N=2 supersymmetric SU(2), the electromagnetic vector superfield
$A_\mu^a$ has a scalar partner $\varphi$ in the adjoint (obtained by
succesive application of the two supersymmetries). We can construct a
gauge invariant complex scalar field, $u={\rm Tr}\,\varphi^2$, which
parametrizes inequivalent vacua, see Ref.~\cite{review}, and can be
geometrically associated with a complex moduli space. It is this
scalar condensate $u$ which plays the role of the inflaton, once N=2
is softly broken to N=0 and the scalar potential acquires a minimum.
In the Seiberg-Witten solution~\cite{SW} one can distinguish two
regions, the perturbative or weak-coupling Higgs region and the
confining monopole/dyon region of strong coupling. In our model, the
soft breaking of N=2 supersymmetry will give mass only to the
monopole, and the dyon region will disappear~\cite{Luis}. Thanks to
duality we will be able to consistently describe our low energy
effective action in terms of the local degrees of freedom of the
corresponding region. In particular, in the region where the monopole
acquires a vacuum expectation value (VEV) the light degrees of freedom
are the monopoles while the `electrons' are confined into electric
singlets. This will play an important role in reheating in this
model, see~\cite{JFL}.

The dynamical scale $\Lambda$ of SU(2) is the only free parameter of
the Seiberg-Witten solution, and can be related via $\Lambda = M
\exp(iS)$ to the dilaton spurion superfield $S=(\pi/2) \tau +
\theta^2F_0$ that breaks supersymmetry, where $\tau = 4\pi i/g^2 +
\theta/2\pi$ is the generalized gauge coupling, see~\cite{Luis}. Here
$M \equiv M_{\rm P}/\sqrt{8\pi}$ is the reduced Planck mass, which
sets the fundamental scale in the problem. In our model we also have
$f_0=\langle F_0\rangle$, the soft supersymmetry breaking parameter,
which should be smaller than $\Lambda$ in order to use the exact
results of Seiberg-Witten. This exact solution is the first term in
the expansion of the low-energy effective action, which because of
supersymmetry can be related to an expansion in $f_0$, to order
$f_0^2$.

The {\rm exact} N=0 low-energy effective scalar potential satisfying
the analyticity and duality properties of the Seiberg-Witten solution
can be written as~\cite{Luis}
\begin{eqnarray}\label{Vm}
V(u) &=& - {2\over b_{11}}\,\rho^4
- {{\rm det}\,b\over b_{11}} \, f_0^2\,, \\
\rho^2 &=& - b_{11} |a|^2 + {f_0\over\sqrt2}|b_{01}|\,,
\hspace{5mm} (\rho>0)
\end{eqnarray}
in terms of the coordinates in the monopole region,
\begin{eqnarray}\label{taum}
a(u) &=& {4i\Lambda\over\pi}\,{E'-K'\over k}\,, \hspace{8mm}
\tau_{11} = {i K\over K'}\,, \\
\tau_{01} &=& {2i\Lambda\over kK'}\,, \hspace{8mm}
\tau_{00} = {8i\Lambda^2\over\pi}\,
\left({E'\over k^2K'}-{1\over2}\right)\,.
\end{eqnarray}
The functions $K(k)$ and $E(k)$ are complete elliptic functions of the
first and second kind respectively, and $E'(k)\equiv E(k')$, where
$k'^2 + k^2 = 1$. All these are functions of $k^2=2/(1+u/\Lambda^2)$
in the complex moduli plane $u$. The functions $b_{ij}\equiv
{\rm Im}\,\tau_{ij}/4\pi$. The scalar potential is thus a non-trivial
function of the inflaton field.

In order to study the cosmological evolution of the inflaton 
under the potential (\ref{Vm}) one should embed this model in
supergravity. This was also performed in Ref.~\cite{Luis} and found
that the only gravitational corrections to the scalar potential were
proportional to $m_{3/2}^2f_0^2$ and therefore completely negligible
in our case, since inflation in this model turns out to occur at the
GUT scale, see below, and thus much above the gravitino mass scale.

One has to take into account however the non-trivial K\"ahler
metric for $u$~\cite{Luis}
\begin{eqnarray}\label{metric}
ds^2 &=& {\rm Im}\,\tau_{11} da\,d\bar a = 
{\rm Im}\,\tau_{11} \left|{da\over du}\right|^2 du\,d\bar u 
\nonumber\\ &\equiv& {1\over2}{\cal K}(u)\,du\,d\bar u \,,
\end{eqnarray}
where ${\cal K}(u)/2$ is the K\"ahler metric. The Lagrangian for the
scalar field $u$ in a curved background can then be written as
\begin{eqnarray}\label{Lagrange}
{\cal L} &=& {1\over2}{\cal K}(u)\,g^{\mu\nu}\partial_\mu u\,
\partial_\nu\bar u - V(u)\,,\\
&=& {1\over2} g^{\mu\nu}\partial_\mu \phi\,
\partial_\nu\bar \phi - V(\phi)\,,
\end{eqnarray}
where $g_{\mu\nu}$ is the spacetime metric, and we have redefined the
inflaton field through $d\phi \equiv {\cal K}(u)^{1/2} du$. We have
plotted in Fig.~1 the scalar potential $V(\phi)$ as a function of the
real and imaginary parts of $\phi$. We have added a constant to the
potential in order to ensure that the absolute minimum is at zero
cosmological constant. In the absence of a working mechanism for the
vanishing of this constant, we have to fix it by hand. Fortunately we
can safely do this without affecting the consistency of the theory
because we have started with a softly broken supersymmetric model.

\begin{figure}[t]
\centering
\hspace*{-4mm}
\leavevmode\epsfysize=6.35cm \epsfbox{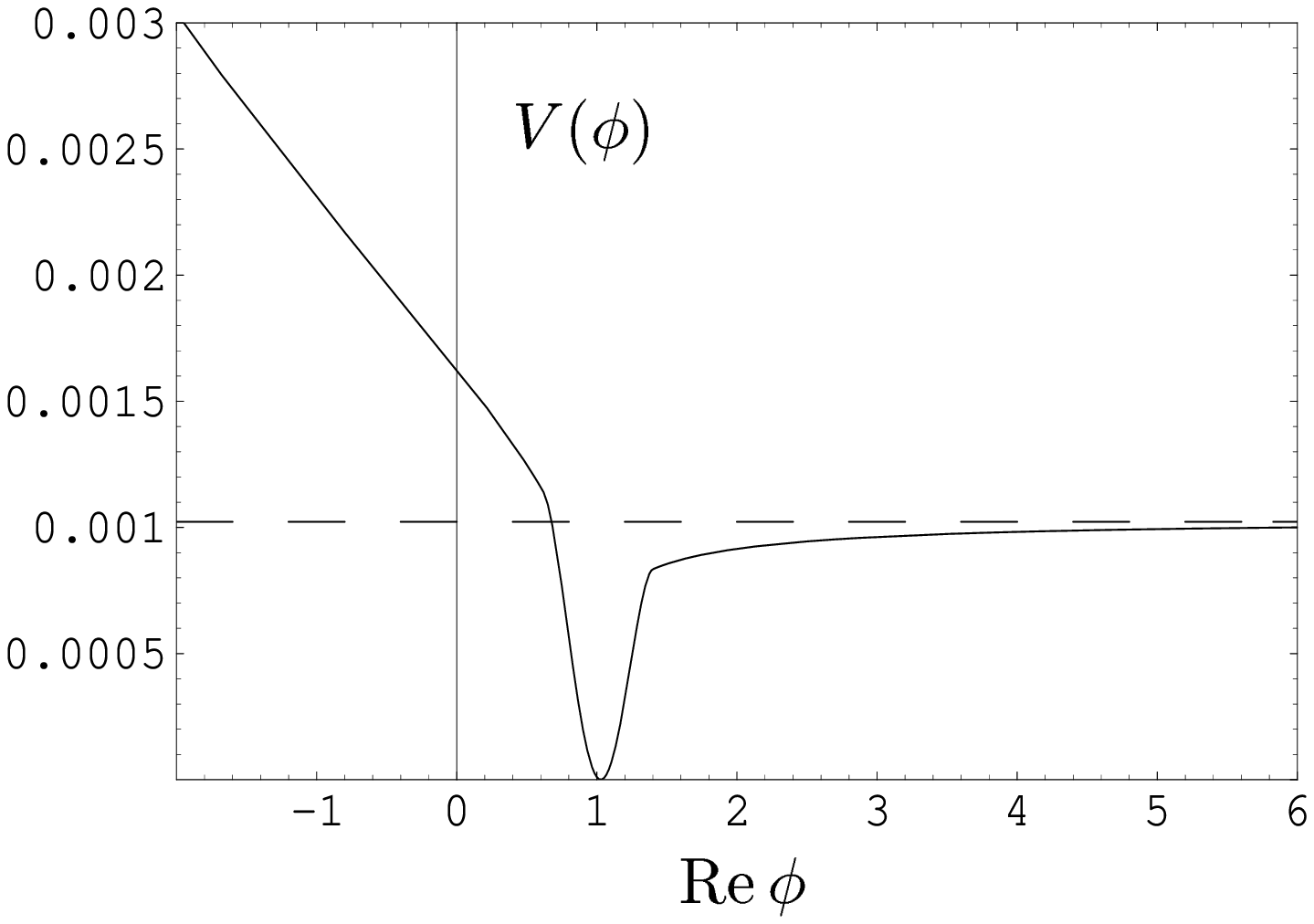}\\[1mm]
\hspace*{-4mm}
\leavevmode\epsfysize=6.35cm \epsfbox{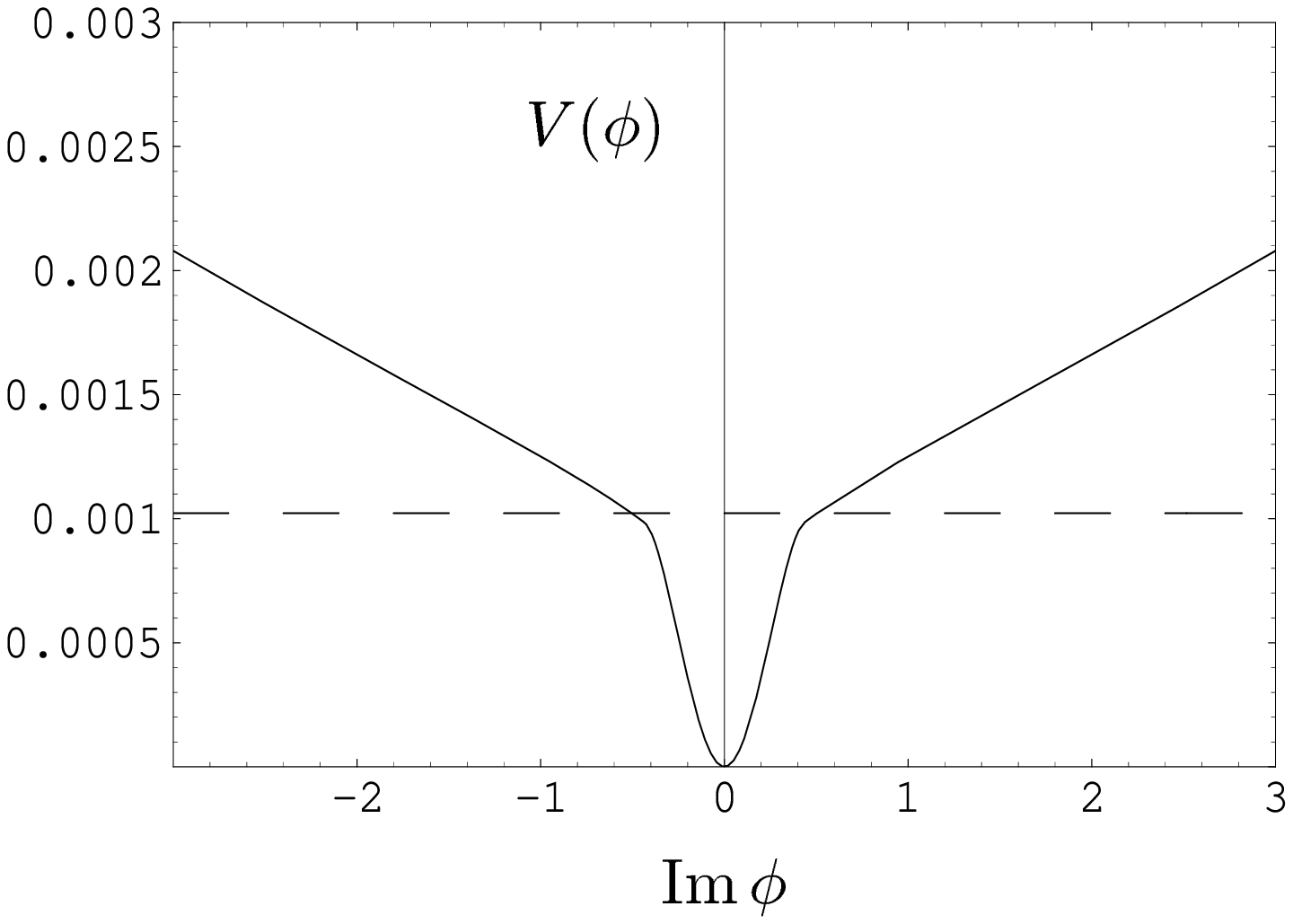}\\[3mm]
\caption[fig1]{\label{fig1} The exact scalar potential $V(\phi)$ in
  moduli space. The top panel shows the potential as a function of
  ${\rm Re}\,\phi$ for ${\rm Im}\,\phi=0$ and presents a minimum at
  $\phi_{\rm min} = \Lambda + f_0/\sqrt8$. The lower panel shows the
  potential as a function of ${\rm Im}\,\phi$ for ${\rm Re}\,\phi =
  \phi_{\rm min}$. The plateau (dashed line) has a vacuum energy
  density of $V_0 = f_0^2\Lambda^2/\pi^2$. We have plotted the figure
  in units of $\Lambda=1$, for $f_0 = 0.1\Lambda$.}
\end{figure}

The flatness of this potential at ${\rm Im}\,\phi =0, {\rm Re}\,\phi >
\phi_{\rm min}$ looks like an excellent candidate for inflation. It
was not included by hand but arised naturally from the soft-breaking
of supersymmetry~\cite{Luis}, albeit with a complicated functional
form (\ref{Vm}). There are two parameters in this model, the dynamical
scale $\Lambda$ and the symmetry breaking scale $f_0$.  Non of these
have to be fine tuned to be small in order to have successful
inflation.  Moreover, as we will show, the trajectories away from the
positive real axis do not give inflation, which simplifies our
analysis significantly. We will now consider the range of values of
$f_0$ and $\Lambda$ that give a phenomenologically viable model.

Let us write down the classical equations of
motion for the homogeneous field $\phi\equiv{\rm Re}\phi$,
\begin{eqnarray}\label{EQM}
H^2 &=& {1\over3M^2}\,\Big[
{1\over2}\dot\phi^2 + V(\phi)\Big]\,, \\
\dot H &=& - {1\over2M^2}\,\dot\phi^2 \,, \\
\ddot\phi &+& 3H\dot\phi + V'(\phi) = 0\,.
\end{eqnarray}
Inflation will be ensured as long as $\epsilon \equiv -\dot H/H^2 < 1$.
The end of inflation occurs at the value of $\phi$ for which
$\dot\phi^2 = V(\phi)$. We can now apply the usual machinery to study
inflationary cosmology with a scalar field potential~\cite{book,LL93}.
It turns out that for all values of the parameters $\Lambda$ and
$f_0<\Lambda$, the extreme flatness of the potential at $\phi>
\phi_{\rm min}$ allows one to use the slow roll approximation in the Higgs
region all the way to the monopole region, where the slope of the
potential is so large that inflation ends and reheating starts as the
condensate oscillates around the minimum. In the slow roll
approximation we can construct the dimensionless parameters~\cite{LL93}
\begin{eqnarray}\label{slowroll}
\epsilon &=& {M^2\over2}\left({V'(\phi)\over V(\phi)}\right)^2\,, \\
\eta &=& M^2\,{V''(\phi)\over V(\phi)} \,,\\ \label{N}
N &=& {1\over M^2} \int\,{V(\phi)\,d\phi\over V'(\phi)}\,,
\end{eqnarray}
where $N$ is the number of $e$-folds to the end of inflation.

In order to know what is the (approximately constant) rate of
expansion during inflation, one has to find the value of the potential
at the minimum, $V_0\equiv |V(u_{\rm min})|$, see Eq.~(\ref{Vm}),
as a function of the symmetry breaking parameter $f_0$. It can be
shown, based on monodromy arguments~\cite{review}, that in the
monopole region close to the minimum, $a(u)\propto (u-\Lambda^2)$.
This ensures that the deviation from $\Lambda^2$ of the position of
the minimum $u_{\rm min}$ is linear in $f_0$, which then implies, to
very good accuracy, that the minimum of the potential is quadratic in
$f_0$,
\begin{eqnarray}\label{minimum}
u_{\rm min} &=& \Lambda^2\Big(1 + {f_0\over2\sqrt2\Lambda}\Big)\,, 
\\ \label{V0}
V_0 &=& {f_0^2\Lambda^2\over\pi^2}\Big(1+{f_0\over8\sqrt2\Lambda}\Big)
\simeq {f_0^2\Lambda^2\over\pi^2} \,.
\end{eqnarray}
These expressions are excellent approximations in the range of values
of $0\leq f_0\leq0.5$, see Fig.~1.

In the Higgs region, along the positive real axis, it is possible to
write in a compact way the K\"ahler metric and the scalar potential
\begin{eqnarray}\label{kahler}
{\cal K}(u)&=&{2k^2\over\Lambda^2\pi^2}\,K K' \,,\\ \label{poten}
V(u)&=&{f_0^2\Lambda^2\over\pi^2}\left[1 - 2\Big({K-E\over k^2K}-
{1\over2}\Big)\right]\,.
\end{eqnarray}

We can now address the issue of what are the parameters $f_0$ and
$\Lambda$ which ensure at least 60 $e$-folds of inflation in order to
solve the classical problems of Big Bang cosmology, as well as to
produce the correct amount of metric fluctuations required to explain
the observed temperature anisotropies of the CMB.

As a consequence of the factorization of the symmetry breaking
parameter $f_0^2$ in the potential (\ref{poten}), the slow-roll
parameters (\ref{slowroll})--(\ref{N}) do not depend on $f_0$. This
further simplifies our analysis. For a given value of $\Lambda/M$ it
is easy to find the value of $\phi_e$ at the end of inflation and from
there compute the value $\phi_{60}$ corresponding to $N=60$ $e$-folds
from the end of inflation.

Quantum fluctuations of the scalar condensate, $\delta \phi$, will
create perturbations in the metric, ${\cal R} = H\delta\phi/\dot\phi$,
which cross the Hubble scale during inflation and later re-enter
during the matter era. Those fluctuations corresponding to the scale
of the present horizon left 60 $e$-folds before the end of inflation,
and are responsible via the Sachs-Wolfe effect for the observed
temperature anisotropies in the microwave background~\cite{COBE}. From
the amplitude and spectral tilt of these temperature fluctuations we
can constrain the values of the parameters $\Lambda$ and $f_0$.

Present observations of the power spectrum of temperature anisotropies
on various scales, from COBE DMR to Saskatoon and CAT experiments,
impose the following constraints on the amplitude of the tenth
multipole and the tilt of the spectrum~\cite{Charley},
\begin{eqnarray}\label{ampli}
Q_{10} &=& 17.5\pm1.1\ \mu{\rm K}\,,\\
n &=& 0.91 \pm 0.10\,.\label{tilt}
\end{eqnarray}
Assuming that the dominant contribution to the CMB anisotropies comes form
the scalar metric perturbations, we can write~\cite{Charley} 
$A_S = 5\times10^{-5}\,(Q_{10}/17.6\ \mu{\rm K})$, where~\cite{LL93}
\begin{eqnarray}\nonumber
A_S^2 &=& {1\over M^2}\Big({H\over2\pi}\Big)^2{1\over2\epsilon}\,,\\[1mm]
n &=& 1 + 2\eta - 6\epsilon\,,\label{PRN}
\end{eqnarray}
in the slow-roll approximation. 

Since during inflation at large values of $\phi$, corresponding to
$N=60$, the rate of expansion is dominated by the vacuum energy
density~(\ref{V0}), we can write, to very good approximation,
$H^2=f_0^2\Lambda^2/3\pi^2M^2$, and thus the amplitude of scalar
metric perturbations is
\begin{eqnarray}\label{PR}
A_S^2 = {1\over24\pi^4}{f_0^2\Lambda^2\over M^4}
{1\over\epsilon_{60}}\,.
\end{eqnarray}

Let us consider, for example, a model with $\Lambda=0.1M$. In that
case the end of inflation occurs at $\phi_e\simeq1.5\Lambda$, still in
the Higgs region, and 60 $e$-folds correspond to a relatively large
value, $\phi_{60} = 14\Lambda$, deep in the weak coupling region. The
corresponding values of the slow roll parameters are $\epsilon_{60} =
2\times10^{-5}$ and $\eta_{60} = -0.04$, which gives $A_S^2 =
5f_0^2/24\pi^4\Lambda^2$ and $n=0.91$. In order to satisfy the
constraint on the amplitude of perturbations~(\ref{ampli}), we require
$f_0=10^{-3}\Lambda$, which is a very natural value from the point of
view of the consistency of the theory.  In particular, these
parameters correspond to a vacuum mass scale of order
$V_0^{1/4}\simeq4\times10^{15}$\,GeV, very close to the GUT scale. For
other values of $\Lambda$ we find numerically the relation
$\log_{10}(f_0/\Lambda) = -4.5 + 1.54\,\log_{10} (M/\Lambda)$, which
is a very good fit in the range $1\leq M/\Lambda \leq 10^3$. For
$M>800\Lambda$, the soft breaking parameter $f_0$ becomes greater than
$\Lambda$, where our approximations breaks down, and we can no longer
trust our exact solution. Meanwhile the spectral tilt is essentially
invariant, $n=0.913 - 0.003\,\log_{10} (M/\Lambda)$, in the whole
range of $\Lambda$. It is therefore a concrete prediction of the
model. Surprisingly enough it precisely corresponds to the observed
value~(\ref{tilt}). This might change however when future satellite
missions will determine the spectral index $n$ with better than 1\%
accuracy~\cite{Kamion}.

There are also tensor (gravitational waves) metric perturbations in
this model, with amplitude $A_T^2 = 2H^2/\pi^2M^2 =
2f_0^2\Lambda^2/3\pi^4M^4$ and tilt $n_T = -2\epsilon$, see
Ref.~\cite{LL93}. The relative contribution of tensor to scalar
perturbations in the microwave background on large scales can be
parametrized by $T/S\simeq12.4\epsilon$. A very good fit to the ratio
$T/S$ in this model is given by $\,\log_{10}(T/S) = -2.6 -
0.91\,\log_{10} (M/\Lambda)$, in the same range as above. Since
$M\geq\Lambda$, we can be sure that no significant contribution to the
CMB temperature anisotropies will arise from gravitational waves.

We still have to make sure that motion along the transverse direction
(for ${\rm Im}\,\phi\neq0$) does not lead to inflationary
trajectories, since otherwise our simple one-dimensional analysis
would break down and there could exist isocurvature as well as
adiabatic perturbations. For that purpose we computed the
corresponding epsilon parameter $(-\dot H/H^2)$ along the imaginary
axis and confirmed numerically, for various values of $\Lambda/M$ and
the whole range, $\phi_e < \phi < \phi_{60}$, that $\epsilon({\rm
  Im}\,\phi) > 1$, and thus inflation does not occur there. A similar
result can be found for the epsilon parameter along the negative real
axis. This ensures that whatever initial condition one may have,
eventually the field will end in the inflationary positive real axis.

We have therefore found a new model of inflation, based on {\em exact}
expressions for the scalar potential of a softly broken N=2
supersymmetric SU(2) theory, to all orders in perturbations and with
all non-perturbative effects included. Inflation occurs along the weak
coupling Higgs region where the potential is essentially flat, and
ends when the gauge invariant scalar condensate enters the strong
coupling confining phase, where the monopole acquires a VEV, and
starts to oscillate around the minimum of the potential, reheating the
universe. A simple argument suggests that during reheating explosive
production of particles could occur in this model. The evolution
equation of a generic scalar (or vector) particle has the form of a
Mathieu equation and presents parametric resonance for certain values
of the parameters, see Ref.~\cite{KLS}. Explosive production
corresponds to large values of the ratio $q=g^2\Phi^2/4m^2$, where $g$
is the coupling between $\phi$ and the corresponding scalar field,
$\Phi$ is the amplitude of oscillations of $\phi$, and $m$ is its
mass. As the inflaton field oscillates around $\phi_{\rm min}$ it
couples strongly, $g\sim1$, to the other particles in the
supermultiplet since the minimum is in the strong coupling region. 
The amplitude of oscillations is of order the dynamical scale,
$\Phi\sim\Lambda$, while the masses of all particles (scalars,
fermions and vectors) are of order the supersymmetry breaking scale,
$m\sim f_0\ll\Lambda$, see Ref.~\cite{Luis2}.  This means that the
$q$-parameter is large, thus inducing strong parametric resonance and
explosive particle production~\cite{KLS}. These particles will
eventually decay into ordinary particles, reheating the universe. This
simple picture however requires a detailed investigation, see
Ref.~\cite{JFL}.

We have shown that for very natural values of the parameters it is
possible to obtain the correct amplitude and tilt of scalar metric
perturbations responsible for the observed anisotropies in the
microwave background. Furthermore, in this model, the contribution of
gravitational waves to the CMB anisotropies is very small. Since
inflation occurs only along the positive real axis of moduli space, we
have an effectively one-dimensional problem, which ensures that
perturbations are adiabatic. But perhaps the most interesting aspect
of the model is the fact that the inflaton field is not a fundamental
scalar field, but a condensate, with an exact non-perturbative
effective potential. The most obvious advantage with respect to other
inflationary models based on supersymmetry is that one has control of
the strong coupling reheating phase, thanks to the exact knowledge of
the inflaton potential.

We are assuming throughout that we can embed this inflationary
scenario in a more general theory that contains two sectors, the
inflaton sector, which describes the soft breaking of N=2
supersymmetric SU(2) and is responsible for the observed flatness and
homogeneity of our universe, and a matter sector with the particle
content of the standard model, at a scale much below the inflaton
sector. The construction of more realistic scenarios remains to be
explored, in which the two sectors communicate via some messenger
sector. For example, one could consider this SU(2) as a subgroup of
the hidden $E_8$ of the heterotic string and the visible sector as a
subgroup of the other $E_8$. No-scale supergravity could then be used
as mediator of supersymmetry breaking from the strong coupling
inflaton sector to the weakly coupled visible sector. For the scales
of susy breaking considered above, $f_0\sim 10^{-5} M_{\rm P}$, we can
obtain a phenomenologically reasonable gravitino mass, $m_{3/2} \sim
f_0^3/M_{\rm P}^2 \sim 10$ TeV~\cite{Costas}. This gravity
mediated supersymmetry breaking scenario needs further study, but
suggests that it is possible in principle to do phenomenology with
this novel inflationary model.

\section*{Acknowledgements}

I would like to thank Luis Alvarez-Gaum\'e, Costas Kounnas and
Frederic Zamora for very stimulating discussions.

\


\begin{thebibliography}{99}

\bibitem{book} A. D. Linde, {\it Particle Physics and Inflationary
        Cosmology} (Harwood, Chur, Switzerland, 1990).
      
\bibitem{SW} N. Seiberg and E. Witten, Nucl. Phys. {\bf B426}, 19 
        (1994); N. Seiberg and E. Witten, Nucl. Phys. {\bf B431}, 484 
        (1994).

\bibitem{Luis} L. Alvarez-Gaum\'e, C. Kounnas, J. Distler and
        M. Mari\~no, Int. J. Mod. Phys. A {\bf 11}, 4745 (1996).

\bibitem{JFL} L. Alvarez-Gaum\'e, J. Garc\'\i a-Bellido and F. Zamora, 
        in preparation.

\bibitem{review} L. Alvarez-Gaum\'e and S. F. Hassan, `Introduction
        to S-Duality in N=2 Supersymmetric Gauge Theories', lecture
        notes of the Trieste Summer School (1995), hep-th/9701069;
        A. Bilal, `Duality in N=2 susy SU(2) Yang-Mills theory', 
        lecture notes of the Strasbourg Meeting (1995), hep-th/9601007.

\bibitem{LL93} A. R. Liddle and D. H. Lyth, Phys. Rep. {\bf 231}, 1 (1993).

\bibitem{COBE} C. L. Bennet et al., Astrophys. J. {\bf 464}, L1 (1996).

\bibitem{Charley} C. H. Lineweaver and D. Barbosa, astro-ph/9706077 (1997).

\bibitem{Kamion} G. Jungman, M. Kamionkowski, A. Kosowky and
   D. N. Spergel, Phys. Rev. D {\bf 54}, 1332 (1996).

\bibitem{KLS} L. Kofman, A. Linde and A. Starobinsky, Phys. Rev. Lett.
   {\bf 73}, 3195 (1994); Phys. Rev. D {\bf 56}, 3258 (1997).

\bibitem{Luis2} L. Alvarez-Gaum\'e and
        M. Mari\~no, Int. J. Mod. Phys. A {\bf 12}, 975 (1997).

\bibitem{Costas} C. Kounnas, private communication.

\end{thebibliography}
\end{document}